# Physical encryption and decryption for secure data transmission in optical networks leveraging the temporal Talbot effect and microwave photonics

Chulun Lin,[1,2,†] Taixia Shi,[1,2,†] Yiqing Liu,[1,2] and Yang Chen,[1,2,*]

[1] Shanghai Key Laboratory of Multidimensional Information Processing, School of Communication and Electronic Engineering, East China Normal University, Shanghai, 200241, China
[2] Engineering Center of SHMEC for Space Information and GNSS, School of Communication and Electronic Engineering, East China Normal University, Shanghai, 200241, China
† These authors contributed equally to this paper
* Correspondence to: Y. Chen, ychen@ce.ecnu.edu.cn

**Abstract:** A novel microwave photonic scheme for secure data transmission in optical networks is proposed. The security of the scheme is guaranteed by physical encryption and decryption via the temporal Talbot effect in dispersive mediums. First, the original data is randomized in the digital domain by performing an exclusive OR operation using a random matrix. Subsequently, a time-varying multi-tone electrical signal, which represents the randomized data matrix, is modulated onto an optical carrier. The optical signal after modulation is then phase-modulated by a temporal Talbot array illuminator (TAI) signal, and the optical signal after discrete quadratic phase modulation will lose its original appearance in the frequency domain and be further dispersed in the first dispersive medium. Due to the dispersion that does not match the TAI signal exactly, the waveform after the first dispersive medium is a noise-like signal. Hence, the physical encryption of the original data is successfully achieved. As the optical signal passes a second dispersive medium that makes the total dispersion match the TAI signal, the temporal waveform of the noise-like signal after photodetection is transformed into pulses. "1" and "0" in the randomized data matrix are represented through the presence and absence of pulses, and the physical decryption is achieved. By further processing the recovered data matrix using the random matrix, the original data can be recovered. The physical layer security of the proposed scheme and its fiber transmission capability are demonstrated. 8-Gbit/s data is transmitted, encrypted, and decrypted using two dispersive mediums and an optical fiber of 10 to 200 km, and error-free transmission is achieved. Many factors that affect the encryption, decryption, and transmission performance of the system have been analyzed.

## 1. Introduction

Optical fiber communication is one of the most important data transmission methods in modern society. It plays an important role in both military and civilian fields because of its characteristics of high capacity, low loss, lightweight, and immunity to electromagnetic interference. However, fiber-based data links in optical networks are prone to various security risks. In general, optical fiber cables, particularly single-mode fiber (SMF), are vulnerable to evanescent attacks deliberately inserted at bending points [1–3]. At locations where the optical power is high, such as amplifier outputs, eavesdroppers can evade detection [3]. In such scenarios, preventing unauthorized receivers from obtaining messages in tapped signals becomes critically important, necessitating the implementation of encryption in different layers to ensure information security in optical networks.

Symmetric key encryption is a typical method in which the same key is used for both encrypting and decrypting data. Therefore, both parties involved in encryption and decryption must share this key in advance and keep it confidential. Nevertheless, employing symmetric keys introduces the risk of theft when the key needs to be shared in the optical network. To avoid using the shared key, asymmetric key encryption methods can be implemented. Despite its high level of security, both encryption and decryption using asymmetric keys involve complex mathematical computations, leading to relatively slow speeds [4,5]. Even so, digital encryption algorithms are currently encountering numerous challenges. Reports have indicated significant advancements in quantum computing [6,7], which possess the capability to crack complex keys at speeds far surpassing that of conventional computers.

To address the security challenges in optical networks posed by the limitations of digital encryption algorithms, physical layer encryption in optical fiber communication is very important for data security. Physical layer encryption can be achieved by quantum key distribution (QKD). QKD utilizes the inimitability of unknown quantum states, making it impossible to access the transmitted information in the absence of a quantum key [8,9]. The security of QKD is guaranteed since the quantum key in QKD eliminates the need for sharing as usual. The possible QKD applications are also investigated [10–14]. However, due to quantum states being irreproducible, the implementation of a repeater in the traditional sense of communications engineering becomes unfeasible. As a result, the limitation inherent in quantum states significantly constrains their practical utilization in optical networks. Chaotic optical fiber communication has been a highly popular physical layer encryption approach in the past few years [15]. In this kind of approach, a wideband chaotic optical signal is used to carry the data, resulting in the generation of a noise-like transmitted optical signal. At the receiver, this noise-like optical signal is demodulated via chaos synchronization [16–18]. However, in the signal recovery process, to extract the chaotic carrier from the received optical signal, the chaotic signal between the transmitter and receiver should be strictly synchronized in the optical network. Meanwhile, the dispersion-compensating fibers for offsetting dispersion over long-distance optical fiber communication are necessary for optical chaos encryption, making the optical network complex and expensive. Therefore, it is imperative to explore potential solutions that avoid the above issues.

Dispersion is an inherent characteristic of optical fiber, and in an optical network, optical fiber is the main medium for signal distribution. Commonly, dispersion can distort optical signals in the temporal domain, which is harmful to optical fiber communication. However, also owing to its ability to distort signals, numerous applications employing dispersion as a signal processor are developed and implemented to realize some special functions, for example, passive amplification [19], ultrafast optical Fourier transformation [20], and optical time-mapped spectrograms [21–23]. In these applications, the basic principles employed are similar: A periodic temporal quadratic phase is loaded on the optical signal, and by matching the temporal quadratic phase and the dispersion, the waveform dispersed over a large time scale can be compressed into narrow pulses in the temporal domain. However, when the dispersion is not accurately matched with the temporal quadratic phase, the optical signal is not compressed and shows a noise-like waveform. Only when the dispersion precisely matches the temporal quadratic phase can a series of narrow pulses be compressed in the temporal domain.

Therefore, if we can encode the data onto the optical signal before undergoing quadratic phase modulation, and subsequently apply the quadratic phase modulation along with a mismatched dispersion to noise the signal in the temporal domain, the data for transmission in the optical network can be hidden within the noise-like signal. At the receiving end, by further matching the dispersion with another dispersive medium (DM), a series of time-domain pulses can be restored. If there exists a mapping relationship between these time-domain pulses and the data encoded onto the optical signal, secure data transmission can be achieved. This phenomenon shares similarities with the utilization of multimode fiber scattering in the context of information security [24]. Considering that optical fiber is the propagation medium in optical networks, there is potential for extending physical layer security research in optical fiber communication based on dispersion.

In this article, based on the above idea, a microwave photonic secure data transmission scheme for optical networks is proposed. The confidentiality of the scheme is assured by physical encryption and decryption via the temporal Talbot effect in dispersive fibers. Initially, at the transmitting end of the optical network, an exclusive OR (XOR) operation is performed on the original data using a random matrix to achieve randomization of data in the digital domain. Subsequently, a time-varying multi-tone electrical signal representing the randomized data matrix is modulated onto an optical carrier. The optical signal after modulation is then phase-modulated by a temporal Talbot array illuminator (TAI) signal, and the optical signal after phase modulation will lose its original appearance in the frequency domain and be further dispersed in the first DM. Due to the dispersion that does not precisely align with the TAI signal, the waveform undergoes a transformation into a noise-like signal after passing through the first DM. Physical encryption of the original data is thus successfully achieved. When processed by a second DM at the receiving end in the optical network, the noise-like signal can be converted into a sequence of electrical pulses after photodetection. "1" and "0" in the randomized data matrix are represented through the presence and absence of pulses, and the physical decryption is achieved. By further processing the recovered data matrix using the random matrix, the original data can be recovered. The physical layer security of the proposed scheme and its fiber transmission capability are demonstrated. 8-Gbit/s data is transmitted, encrypted, and decrypted using two DMs and an optical fiber of 10 to 200 km, and error-free transmission is achieved. Many factors that affect the encryption, decryption, and transmission performance of the system have been analyzed. Another key advantage of the microwave photonic secure data transmission scheme is that, dispersion, which traditionally acts as a negative factor, is incorporated into the secure data transmission scheme, eliminating the need for dispersion compensation in the optical networks and requiring only dispersion matching through a large DM at the receiving end.

## 2. Principle

*2.1 System structure and principle for secure data transmission in optical fibers*

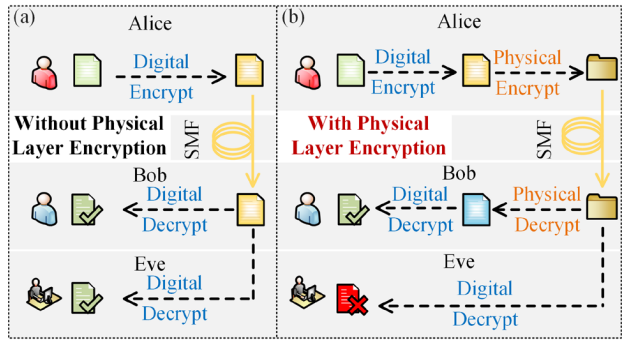

Fig. 1. Data transmission in optical fiber (a) without physical layer encryption (b) with physical layer encryption.

In a typical optical network, an optical fiber communication link commonly only employs digital encryption algorithms. As shown in Fig. 1(a), Alice applies digital encryption and then transmits the encrypted data in the optical network over a section of SMF to Bob. Using the key shared by Alice, Bob receives the optical signal from Alice and decrypts the signal to get the original data transmitted by Alice. Meanwhile, when Eve intercepts the signal, the data from Alice can also be decrypted with the stolen or cracked key. Fig. 1(b) illustrates the underlying principle of the proposed secure data transmission scheme for optical networks with both digital encryption and physical layer encryption. After digital encryption, physical layer encryption is also employed. Hence, the data transmitted in the optical network is double-encrypted. After receiving the optical signal from Alice, Bob still can obtain the original data transmitted by Alice when the physical layer decryption and digital decryption are both implemented. However, Eve, this time cannot obtain the original data from Alice by simply intercepting the signal with the stolen or cracked key and without the physical decryption rule. It is the critical role of physical layer encryption that makes the data transmission, as well as the optical network, more secure.

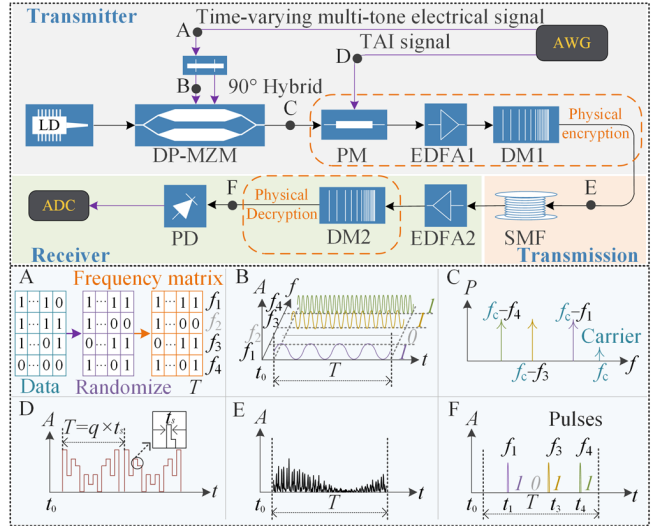

Fig. 2. Proposed secure data transmission scheme for optical networks with physical layer encryption and decryption based on the temporal Talbot effect. LD, laser diode; DP-MZM, dual-parallel Mach–Zehnder modulator; PM, phase modulator; EDFA, erbium-doped fiber amplifier; DM, dispersive medium; AWG, arbitrary waveform generator; SMF, single-mode fiber; PD, photodetector; ADC, analog-to-digital converter. A–F shows the diagrams of the corresponding points in the figure.

Fig. 2 depicts the proposed secure data transmission scheme for optical networks with physical layer encryption and decryption based on the temporal Talbot effect. At a transmitting end in an optical network, a continuous-wave optical signal emitted by a laser diode (LD) serves as the optical carrier and is injected into a dual-parallel Mach–Zehnder modulator (DP-MZM). A time-varying multi-tone electrical signal is generated by an arbitrary waveform generator (AWG) and is applied to the DP-MZM via a 90° electrical hybrid coupler. The frequency of the time-varying multi-tone electrical signal changes once every time interval $T$, and the presence or absence of a series of $N$ different frequencies with a frequency interval of $\Delta f$ within $T$ represents the $N$-bit data transmitted during that time interval $T$. The data stream loaded on the time-varying multi-tone electrical signal is generated after the original data is randomized and further mapped to a frequency matrix. As shown at point A in Fig. 2, the original data is converted to a data matrix and then randomized by a random matrix, so a randomized data matrix with "0" or "1" elements is obtained. The randomized data matrix is mapped to a frequency matrix, where each column represents a distinct time interval $T$, and each row corresponds to a unique signal frequency among $N$ frequencies. Within this matrix, "0" signifies the absence of a frequency, whereas "1" indicates its presence. According to the frequency matrix, the time-varying multi-tone electrical signal is generated, as illustrated by the waveform shown at point B in Fig. 2, in which $N = 4$ is used. Here, the DP-MZM is used as a carrier-suppressed single-sideband (CS-SSB) modulator. The optical signal spectrum after CS-SSB modulation is schematically shown at point C in Fig. 2, where only three negative first-order optical sidebands are preserved according to the signal at point B in Fig. 2. Subsequently, the optical signal from the DP-MZM is sent to a phase modulator (PM), where it is modulated by a specially designed TAI signal also generated from the AWG. The TAI signal is a discrete quadratic signal, which has a period of $T$ and encompasses $q$ distinct amplitude values, each lasting for a duration of $t_s$ [25], as illustrated at point D in Fig. 2. The TAI signal will introduce a discrete quadratic phase to the optical signal.

The phase-modulated optical signal from the PM is amplified by an erbium-doped fiber amplifier (EDFA1) and then dispersed in DM1. To achieve large dispersion and low insertion loss, a linearly chirped fiber Bragg grating with large dispersion can be used as DM1. The TAI signal and the dispersion characteristics of DM1 are designed to be mismatched, so the temporal waveform from DM1 is a noise-like waveform, which is schematically illustrated at point E in Fig. 2. When comparing the output signal of the DP-MZM to the signal after it has passed through the PM and DM1, it becomes evident that the original characteristics in both the time and frequency domains have been significantly altered, losing the original appearance. It becomes infeasible to deduce the concealed data solely by examining either the temporal domain or the frequency domain. Therefore, the physical layer encryption is achieved.

Then the optical signal from DM1 is transmitted in the optical network through a section of SMF to the receiving end, in which the received signal is amplified by EDFA2 and further processed by DM2. When the total dispersion of DM1 combined with DM2 and the SMF matches the parameters of the TAI signal, that is, when the fractional Talbot condition is met, the noise-like temporal optical waveform is compressed into narrow optical pulses in the temporal domain. Thus, the physical layer decryption is achieved. The optical pulses from DM2 are converted to electrical pulses in a photodetector (PD). If all $N$ equally spaced frequencies are present within a given period $T$, it will result in the generation of $N$ equally spaced narrow pulses during that period. Conversely, if a frequency is absent, its corresponding pulse will fail to be produced. The generated electrical pulses are captured by an analog-to-digital converter (ADC) and further post-processed to recover the original data.

As discussed above, the presence or absence of each of $N$ equally spaced narrow pulses in a given period $T$ determines the $N$-bit data transmitted in this period. Due to the influence of noise, in signal processing, the period $T$ is divided into $N$ equal segments, and the signal energy is integrated for each segment of $T/N$, with an appropriate threshold set. When the energy exceeds the threshold, the data transmitted in this segment is considered to be "1", otherwise it is considered to be "0". As shown at point F in Fig. 2, because $f_2$ does not exist in this period, only three pulses representing $f_1$, $f_3$, and $f_4$ are generated. When the data represented by the $N$ segments is determined, the $N$-bit data in this period is recovered. Through the recovery of data in multiple different periods $T$, the randomized data matrix at the transmitting end can be obtained. By performing an XOR operation on the obtained randomized matrix using the same random matrix as is used by Alice at the transmitting end, the original data matrix, and thus the original data, can be recovered by Bob. In comparison, Eve would never obtain the pulse from the temporal waveform without using a proper dispersion provided by DM2. Therefore, the proposed scheme enables secure data transmission in optical networks through the physical layer encryption and decryption.

*2.2 Principle of physical encryption and decryption based on the temporal Talbot effect*

In the proposed scheme, the key to the physical layer encryption and decryption for secure data transmission in optical networks is the temporal Talbot effect. The Talbot effect is first observed in space optics [26]. The diffraction of light in space enables the formation of precise replicas of periodic objects at specific distances away from the object when a beam of coherent light is reflected from the periodic object, and this effect is also called the spatial Talbot self-imaging effect. Based on the Talbot effect, a uniform spatial optical wavefront can be converted into an array of bright light spots, which is known as TAI [27]. Because of the analogy between the diffraction in space and the group velocity dispersion in the temporal domain, the phenomena of spatial Talbot self-imaging effect can also be observed in the temporal domain. The temporal counterpart of the Talbot effect is referred to as the temporal Talbot self-imaging effect [28,29]. By imitating the principle of TAI in space, the process of converting an incoming waveform into a series of short temporal pulses is called temporal TAI [19]. In the proposed scheme, temporal TAI is the most important component and we use the TAI design [25] that is achieved by a CW-to-pulse conversion via a TAI signal. The TAI structure mainly consists of a CW laser source, a PM, a discrete quadratic TAI signal, and a DM. Then, the TAI signal and its relationship with the dispersion value of the DM are briefly discussed and given [19].

The period of the TAI signal is set to $T$. Within each period, the quadratic waveform is divided into $q$ levels and the width of each level is $t_s$. This relationship can be mathematically expressed as

$$T = qt_s. \tag{1}$$

The TAI phase after phase modulation by the TAI signal is designed based on the Talbot effect, according to

$$\varphi_n = -\pi \frac{q-1}{q} n^2, \tag{2}$$

where $n = 1, 2, ..., q$. Assuming the dispersions offered by DM1, the SMF, and DM2 are $\ddot{\Phi}_1$, $\ddot{\Phi}_2$, and $\ddot{\Phi}_3$, respectively, the total dispersion from the transmitting end to the receiving end in the optical network is represented by

$$\ddot{\Phi} = \ddot{\Phi}_1 + \ddot{\Phi}_2 + \ddot{\Phi}_3. \tag{3}$$

To successfully decrypt the received signal, the total dispersion and the TAI phase should satisfy the following condition

$$2\pi\ddot{\Phi} = qt_s^2. \tag{4}$$

For the SMF, $\ddot{\Phi}_2 = |\beta_2|L$, where $|\beta_2|$ is the dispersion coefficient and $L$ is the length of the SMF. $|\beta_2|$ can be further written as

$$|\beta_2| = \frac{\lambda_{ref}^2}{2\pi c} D_\lambda, \tag{5}$$

where $c$ represents the light speed in a vacuum, $D_\lambda$ denotes the dispersion parameter of the SMF, and $\lambda_{ref}$ denotes the reference wavelength of SMF. Under these circumstances, the relationship between the dispersion of DM3 and the length of the fiber, the dispersion of DM1, and the TAI phase is given by the following equation,

$$\ddot{\Phi}_3 = \frac{qt_s^2}{2\pi} - \frac{\lambda_{ref}^2 D_\lambda L}{2\pi c} - \ddot{\Phi}_1. \tag{6}$$

Therefore, when the TAI phase determined by $q$ and $t_s$, the dispersion provided by DM1 at the transmitting end, and the fiber transmission length $L$, are determined, at the receiving end of the optical network, the dispersion of DM3 can be set according to Eq. (6) to implement the physical decryption.

In addition, in the proposed scheme, the time-varying multi-tone electrical signal should be converted to electrical pulses at the receiving end for data recovery. The operating bandwidth of the system $B_w$ is determined by $B_w = 1/t_s$ [22], so the maximum number of frequencies $M$ that can be allocated in a period $T$ is

$$M = \max(N) = \left\lfloor \frac{B_w}{\Delta f} \right\rfloor. \tag{7}$$

Therefore, the data rate $R$ of the scheme can be calculated by $R = N/T$, and the maximum data rate $R_{max} = M/T$.

*2.3 Secrecy rate*

The security performance of the proposed scheme is evaluated by the secrecy rate. It is assumed that the data transmitted by Alice is denoted as $M_A$, and the data received by Bob is denoted as $M_B$. The probability that Bob received incorrect data is expressed as

$$\varepsilon_B = \Pr(M_B \neq M_A). \tag{8}$$

Meanwhile, the data intercepted by Eve from Alice is denoted as $M_E$. The probability that Eve received incorrect data can be expressed as follows

$$\varepsilon_E = \Pr(M_E \neq M_A). \tag{9}$$

The channel capacity $C$ with bit-flip probability $\varepsilon$ is given by [30]

$$C = 1 - H_b(\varepsilon), \tag{10}$$

where $H_b$ represents the binary entropy function, which can be expressed as

$$H_b(\varepsilon) = -\varepsilon \log_2(\varepsilon) - (1-\varepsilon) \log_2(1-\varepsilon). \tag{11}$$

The secrecy rate of the proposed scheme is denoted as $R_S$, which can be calculated as the difference between the channel capacities of Bob and Eve

$$R_S = [C_{Bob} - C_{Eve}]^+ = [H_b(\varepsilon_E) - H_b(\varepsilon_B)]^+, \tag{12}$$

where $C_{Bob}$ and $C_{Eve}$ represent the channel capacities of Bob and Eve, respectively.

## 3. Results

*3.1 Secure data transmission in the optical network*

The proposed system is verified at a sampling rate of 160 GSa/s based on the schematic diagram shown in Fig. 2. The output power of the LD is set to 10 mW and the frequency of the LD is fixed at 193.1 THz. To better display the physical encryption and decryption of the proposed scheme, a binary image with 400×400 pixels is used as the information to be transmitted in the optical network. Black and white in the image are represented by "1" and "0", respectively, which are then converted into a 40×4000 data matrix and randomized by a 40×4000 random matrix. The matrix after randomization is mapped to a 40×4000 frequency matrix, which is used for the generation of the time-vary multi-tone electrical signal carrying the data to be transmitted. It is noted that the parameters of the time-vary multi-tone electrical signal, the TAI signal, the SMF, DM1, and DM2 should be set according to the relationship introduced in Section 2.2. For example, when the TAI signal has $q = 100$ levels and a width of each level of $t_s = 25$ ps, the TAI signal period $T = 2.5$ ns and the operating bandwidth of the system $B_w = 40$ GHz. Under these parameters, if the frequency interval is set to $\Delta f = 1$ GHz, we can transmit 40 frequencies, i.e., 40 bits, per 25 ns, and the data rate is $R = 16$ Gbit/s. In the demonstration, the 40 frequencies are set from 0.5 to 39.5 GHz with an interval of 1 GHz. It should be noted that the frequency interval needs to be set reasonably because if it is set too small, it will cause the adjacent electrical pulses of the output to be difficult to distinguish and increase the bit error rate (BER) of the scheme.

The two EDFAs in the scheme are used to provide enough link gain to compensate for the loss of the DP-MZM, the PM, the SMF, and the two DMs, guaranteeing an optical power of around −5.5 dBm at the input of the PD. As discussed in Section 2, the dispersion of the SMF in the optical network is no longer a negative effect but part of the dispersion required for secure data transmission. Therefore, the only negative effect of the SMF on the system is its loss, which can be compensated by the EDFAs. In the following demonstration, the length of the SMF is first set to 10 km. The dispersion parameter of the SMF is $D_\lambda$ = 16 ps/(nm·km) while the insertion loss is 0.2 dB/km. Since the dispersion introduced by SMF is $\ddot{\Phi}_2$ = 204 ps$^2$/rad and the total dispersion of DM1, the SMF, and DM2 is calculated based on Eq. (4) to be $\ddot{\Phi}$ = 9947 ps$^2$/rad, the dispersions of DM1 and DM2 should be $\ddot{\Phi}_1 + \ddot{\Phi}_3$ = 9743 ps$^2$/rad. This amount of dispersion can be randomly allocated to DM1 and DM2 for encryption and decryption, but it is not advisable to allocate the vast majority of dispersion to either one, as this would degrade the encryption performance. In the demonstration, the dispersion ratio of DM1 and DM2 is set to 4:6.

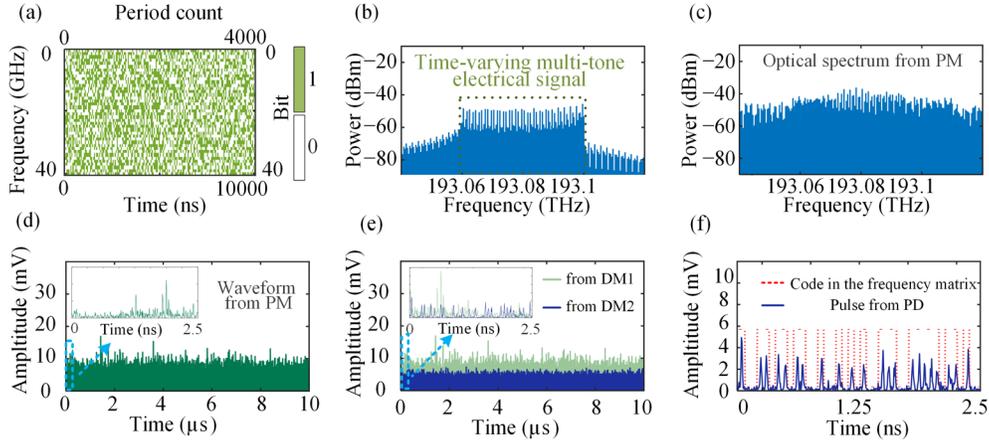

Fig. 3. (a) Randomized frequency matrix. (b) Optical spectrum of the optical signal from the DP-MZM. (c) Optical spectrum of the optical signal from the PM. (d) Waveform of the optical signal from the PM. (e) Waveforms of the optical signal from DM1 and DM2. (f) Electrical pulses from the PD and the corresponding binary data in the randomized data matrix in a period.

By employing the parameters mentioned above, the data, spectrum, and waveform in the proposed secure data transmission scheme in an optical network are investigated and shown in Fig. 3. Fig. 3(a) illustrates the randomized 40×4000 frequency matrix generated from a binary image. According to the matrix, the time-vary multi-tone signal is generated by the AWG and modulated on the optical carrier from the LD at the DP-MZM via CS-SSB modulation. The generated −1st-order optical sideband from the DP-MZM is shown in Fig. 3(b), which is further phase-modulated at the PM by the TAI signal also generated by the AWG. Fig. 3(c) and (d) depict the optical spectrum and the waveform of the optical signal from the PM. As can be seen, after introducing the TAI phase, the optical spectrum is greatly broadened, which makes the spectrum completely different from the original spectrum. At this point, because phase modulation only affects the phase of the signal, the waveform of the signal is still the same as the waveform of the DP-MZM output. Then, DM1 is used to disperse the signal. As shown in Fig. 3(e), after DM1, the waveform is completely different from that shown in Fig. 3(d), which shows a noise-like characteristic. After further dispersed in the SMF of the optical network and DM2 at the receiving end, the waveform from DM2 is also shown in Fig. 3(e). It is observed that the noise-like waveform from DM1 is compressed into multiple distinguished pulses, which carry the data transmitted by the 40×4000 frequency matrix. Then, the optical pulses from DM2 are converted into electrical pulses by the PD and sampled by the ADC. A section of electrical pulses from 0 to 2.5 ns is shown in Fig. 3(f). The binary data transmitted in this period is also given in the figure using a red dotted line. As can be seen, the obtained pulse is highly consistent with the transmitted data.

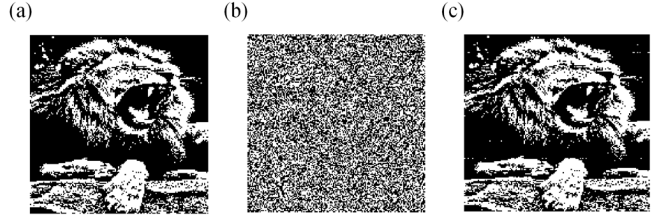

Fig. 4. (a) Binary image transmitted by Alice. (b) Binary image stolen by Eve from the SMF without physical layer decryption by DM2. (c) Binary image recovered by Bob from the SMF with physical layer decryption by DM2.

Fig. 4 shows the results of the binary image transmission in the optical network using the proposed secure data transmission scheme. The binary image transmitted by Alice is a lion, which is shown in Fig. 4(a). After physical encryption, the image is transmitted through the optical network using the proposed secure data transmission scheme. Eve intercepts the optical signal from the SMF. Without physical decryption, the image recovered by Eve is shown in Fig. 4(b) and Eve only gets a confusing image without the assistance of DM2. At the receiving end in the optical network, Bob has DM2 for physical decryption and he can obtain a binary image with the nearly same content as transmitted by Alice, as illustrated in Fig. 4(c). It can be seen that by using the physical layer encryption and decryption, the proposed scheme leaves the data less vulnerable to theft. It should also be noted that the binary image obtained by Bob is slightly different from that transmitted by Alice, i.e., some pixels are demodulated incorrectly.

*3.2 Secrecy rate and BER under different dispersions*

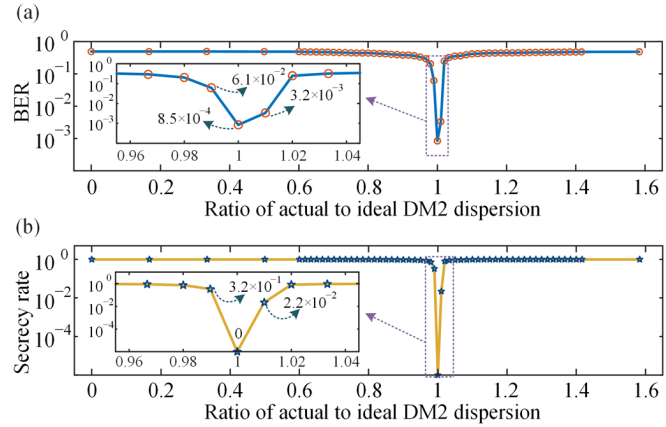

Fig. 5. (a) BER and (b) secrecy rate under different ratios of actual to ideal DM2 dispersion.

Then, the BER and secrecy rate of the proposed scheme are studied under similar parameters. The only difference is the dispersion of DM2, which is changed in this demonstration to show the need for the precise setting of DM2 dispersion. When the dispersion of DM1 is 3897 ps$^2$/rad, the ideal dispersion of DM2 is 5846 ps$^2$/rad. Based on this value, we change the dispersion of DM2. The same 400×400-pixel binary image is transmitted through the proposed scheme, so 1.6×10$^5$ bit is transmitted in one experiment. To get a more accurate BER, the binary image is transmitted in the optical network four times under different noise settings, so a total of 6.4×10$^5$ bits are used to evaluate the BER. The BER with different dispersion of DM2 is given in Fig. 5(a) and the corresponding secrecy rate is given in Fig. 5(b). As shown in Fig. 5, when the dispersion of DM2 is set to the ideal value, i.e., 5846 ps$^2$/rad, the BER reaches the best, which is 8.5×10$^{-4}$. In this case, the secrecy rate is 0, so Bob can get the best data recovery performance using a completely matched DM2. Due to the 8.5×10$^{-4}$ BER, some bits are received with error and this is the reason why the binary image obtained by Bob is slightly different from that transmitted through the optical network by Alice in Fig. 4. When the dispersion of DM2 is 1% less than the ideal value, the BER reaches 6.1×10$^{-2}$ and the secrecy rate changes to 3.1×10$^{-1}$. Similarly, when the dispersion of DM2 is 1% greater than the ideal value, the BER changes to 3.2×10$^{-3}$, and the secrecy rate is 2.2×10$^{-2}$. When the dispersion deviates from the ideal value by 2% or more, the BER is very close to or equal to 0.5 while the secrecy rate is also very close to or equal to 1. The BER and secrecy rates under different dispersions of DM2 indicate that Eve has impossible access to useful information in the absence of a well-matched DM to achieve the physical layer decryption. Meanwhile, it is a costly affair for Eve to decrypt the stolen optical signal through different DMs. Furthermore, a time-varying DM or a time-very TAI signal can be used to completely avoid Eve's cracking of the dispersion value, which will be discussed in Section 4.2.

*3.3 Influence of the number of phase levels in the TAI signal*

Although the proposed scheme allows for the secure transmission of data in optical networks, it remains highly desirable to achieve error-free signal reception at high communication data rates. To investigate the factors affecting the BER of the system, the number of the phase levels $q$ of the TAI signal is first examined. In this study, the data rate is fixed at 16 Gbit/s and $t_s$ is fixed at 25 ps. The number of the phase levels $q$ is set to 40, 80, 100, 160, and 200, while the corresponding frequency interval $\Delta f$ is 2.5 GHz, 1.25 GHz, 1 GHz, 0.625 GHz, and 0.5 GHz. The total dispersion $\ddot{\Phi}$, the TAI signal period $T$, and the number of bits $N$ per TAI signal period are set according to the relationships discussed in Section 2.2. Besides, other parameters in this study are configured as that used in obtaining Fig. 3.

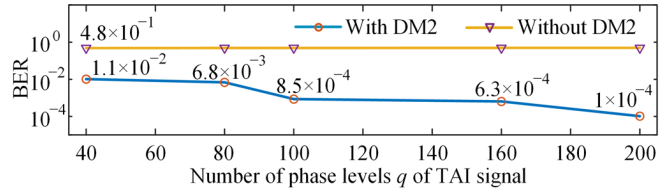

Fig. 6. BER under different numbers of phase levels $q$ of the TAI signal.

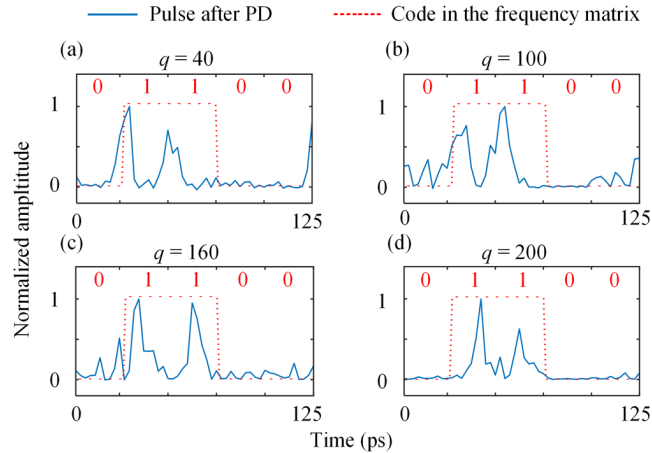

Fig. 7. Temporal waveforms from 0 to 125 ps under different numbers of phase levels $q$ of the TAI signal. (a) $q = 40$, (b) $q = 100$, (c) $q = 160$, (d) $q = 200$.

Fig. 6 shows the BER at the receiving end in the optical network with and without physical layer decryption by DM2 under different $q$. When the data rate is fixed at 16 Gbit/s and the physical decryption is implemented by DM2, a notable improvement in the BER is observed with the increase of $q$ from 40 to 200. Specifically, the BER undergoes a two-order-of-magnitude improvement from $1.1\times10^{-2}$ to a significantly lower level of $1\times10^{-4}$. In comparison, without physical decryption, the BER is always kept at around 0.5 due to the data encryption. The results given in Fig. 6 indicate that increasing the number of phase levels $q$ can significantly enhance the demodulation capability of the receiving end, thus improving the transmission performance of the proposed scheme. The reason is given in Fig. 7, which shows the temporal waveform from the PD under different $q$. Evidently, with the increase in $q$, the width of the pulses generated by the scheme becomes narrower, which reduces the mutual interferences between adjacent pulses. Additionally, as can be seen from Fig. 7, as $q$ increases, the number and amplitude of unwanted interference pulses and spurs are continuously reduced. Therefore, at the same data rate, an increase in $q$, as well as the total dispersion $\ddot{\Phi}$, leads to a significant improvement in BER. When $t_s$ is fixed, increasing $q$ does not lead to an increase in the difficulty of generating the TAI signal or generating the time-varying multi-frequency signal, for the TAI signal bandwidth and time-varying multi-frequency signal are both determined by $t_s$. The cost is a larger amount of dispersion is needed at the receiving end of the optical network.

*3.4 Influence of the width of each level $t_s$ in the TAI signal*

In this study, the influence of the width of each level $t_s$ in the TAI signal on the performance is evaluated. The data rate is fixed at 16 Gbit/s and the number of phase levels $q$ in the TAI signal is fixed at 100. The width of each level $t_s$ in the TAI signal is set to 50 ps, 25 ps, and 12.5 ps, respectively, so the corresponding frequency interval $\Delta f$ is 0.25 GHz, 1 GHz, and 4 GHz. The total dispersion $\ddot{\Phi}$, the TAI signal period $T$, and the number of bits $N$ per TAI signal period are also set according to the relationships discussed in Section 2.2. Besides, other parameters in this study are also configured as that used in obtaining Fig. 3.

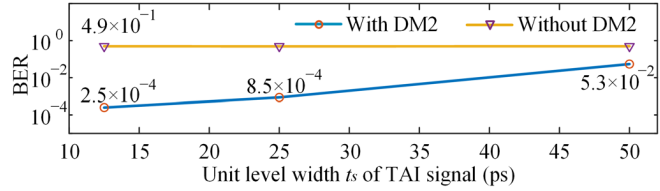

Fig. 8. BER under different $t_s$ of TAI signal.

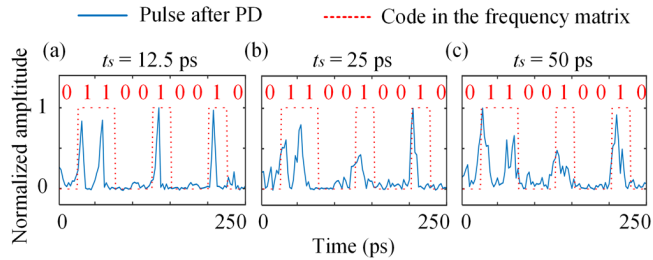

Fig. 9. Temporal waveform from 0 to 250 ps under different $t_s$ of the TAI signal. (a) $t_s$ = 12.5 ps, (b) $t_s$ = 25 ps and (c) $t_s$ = 50 ps.

Fig. 8 shows that when $t_s$ decreases from 50 ps to 12.5 ps, the BER at the receiving end in the optical network using a dispersion-matched DM2 is improved from $5.3\times10^{-2}$ to $2.5\times10^{-4}$ at the same bit rate of 16 Gbit/s, while the BER without physical layer decryption is always kept at around 0.5. The results indicate that decreasing $t_s$ is an effective way to improve the data transmission performance. Fig. 9 shows the temporal waveform from the PD under different $t_s$. Similar to the results presented in Fig. 7, the improvement in BER in Fig. 8 under different $t_s$ is mainly due to the pulses becoming narrower and more regular, as well as the reduced interference and spur. However, unlike the case in Section 3.3 where increasing $q$ improves performance without affecting the complexity of the signals the system needs to generate, reducing $t_s$ will significantly increase the difficulty of system implementation, as the value of $t_s$ directly determines the bandwidth of the TAI signal and the time-frequency multi-frequency signals. Nevertheless, there is also a benefit to this approach: by reducing $t_s$ while maintaining the same $q$, it can significantly decrease the amount of dispersion required at the receiving end in the optical network.

### 3.5 Data transmission under different system configurations

After optimizing the proposed scheme via the TAI signal parameters on the BER, the scheme is further studied under various alternative system configurations. Here, the data rate is set to 8 Gbit/s, 16 Gbit/s, or 32 Gbit/s, $q$ is set to 100 or 200, and $t_s$ is set to 12.5 ps, 25 ps, or 50 ps. 8 groups of system configurations for comparison are given in Table 1.

**Table 1. Different System Configurations**

| System Configurations | $t_s$ | $q$ | Frequency Interval $\Delta f$ | Number of Frequencies $N$ | Data rate |
|---|---|---|---|---|---|
| 1 | 50 ps | 100 | 0.5 GHz | 40 | 8 Gbit/s |
| 2 | 25 ps | 100 | 1 GHz | 40 | 16 Gbit/s |
| 3 | 25 ps | 100 | 2 GHz | 20 | 8 Gbit/s |
| 4 | 25 ps | 200 | 0.5 GHz | 80 | 16 Gbit/s |
| 5 | 25 ps | 100 | 0.5 GHz | 80 | 32 Gbit/s |
| 6 | 12.5 ps | 100 | 2 GHz | 40 | 32 Gbit/s |
| 7 | 12.5 ps | 200 | 2 GHz | 40 | 16 Gbit/s |
| 8 | 12.5 ps | 200 | 1 GHz | 80 | 32 Gbit/s |

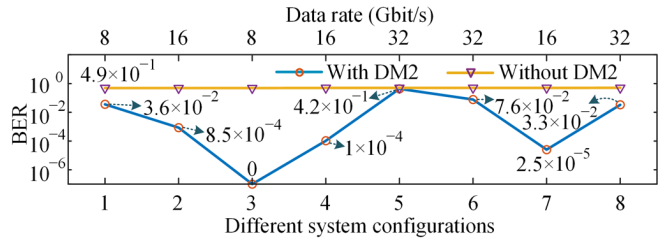

Fig. 10. BER under different system configurations.

Fig. 10 shows the BER of the 8 groups of system configurations with and without physical decryption using DM2. Some configurations have been discussed previously, while others have not. However, as shown in Fig. 10, the conclusions drawn in Sections 3.3 and 3.4 are valid under the same bit rate. It is also indicated from Fig. 10 that, while the conclusions in Sections 3.3 and 3.4 hold true when the data rate is 32 Gbit/s, achieving a very good BER result becomes difficult due to the high data rate. At a data rate of 16 Gbit/s, although the BER can be improved to a $10^{-5}$ level, error-free data recovery is still not achieved. In comparison, when the data rate is 8 Gbit/s, error-free data recovery can be achieved under Configuration 3.

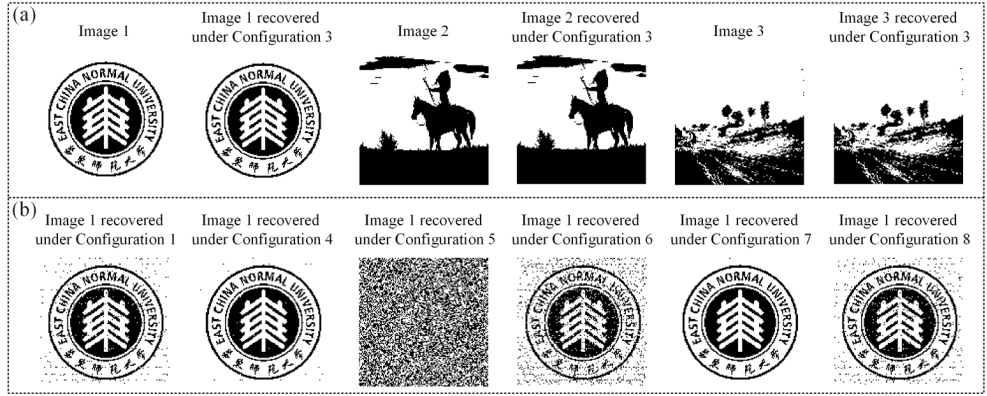

Fig. 11. Binary images transmitted and received by the proposed scheme under different system configurations. (a) Different images under Configuration 3. (b) image 1 under Configurations 1, 4, 5, 6, 7, and 8.

Then three binary images are employed, encrypted, and decrypted using the proposed scheme under different system configurations in Table 1. Fig. 11(a) shows the original binary image and the binary image recovered at the receiving end under error-free Configuration 3. It can be seen that there are no blurred parts or inverted pixels in the recovered images. When other configurations are used, as shown in Fig. 11(b), varying degrees of image blurring occur. In Configuration 5, the original image cannot be identified. Obviously, the difference between the recovered image and the original image under different configurations is determined by the BER shown in Table 1. In fact, although it is difficult for the proposed system to achieve error-free transmission under some configurations, in practical communication systems, error-correction codes with error-correcting capabilities, such as Hamming codes and low-density parity-check codes, would be introduced in the optical network for error control. Under such circumstances, depending on the error-correcting capability of the error-correction codes, many of the configurations in Table 1 would be capable of achieving error-free transmission when combined with these error-correction codes.

*3.6 System performance under non-ideal system parameters*

In the proposed scheme, three factors are very important for the generated pulses carrying the binary data: (1) The modulation index of the PM, (2) The ideality of the TAI signal, and (3) The bandwidth of the PD. The modulation index of the PM should be set to $2\pi$ [25] in the proposed scheme, which is actually a very high modulation index for a modulator and requires the TAI signal applied to the PM to have a very large amplitude. If the modulation index does not reach $2\pi$, the system performance will definitely decrease. Besides, for the TAI signal, due to the very small width of each level, it should have a very large signal bandwidth. However, in practical systems, the signal bandwidth is limited, so we will also evaluate the influence of TAI signal bandwidth on the system performance. In addition, after the optical signal is received at the receiving end, a PD is used to convert the optical pulses to electrical pulses. If the PD bandwidth is insufficient, it will bring the problem of pulse broadening, which will affect the system's performance.

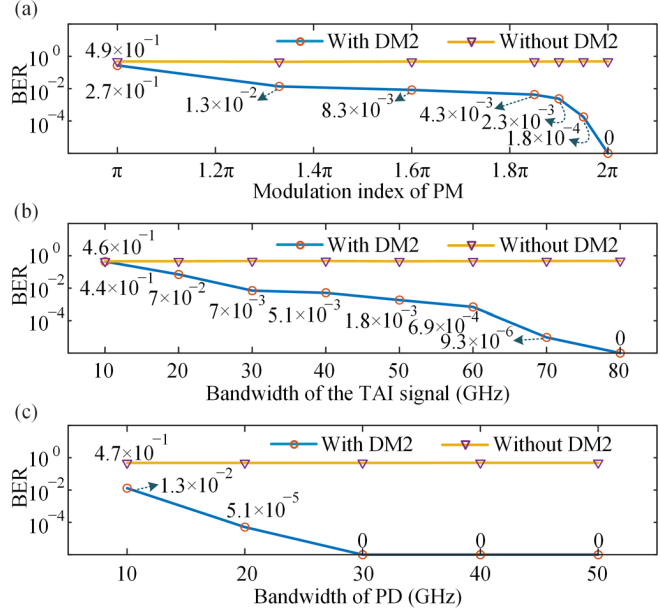

Fig. 12. BER under different (a) modulation indices, (b) bandwidths of the TAI signal, and (c) PD bandwidths.

In this study, based on error-free Configuration 3, the performance of the system is investigated under different non-ideal system parameters by changing the above three parameters separately. When one of the parameters is studied, the other two parameters are set to ideal values. As shown in Fig. 12(a), with the decrease of the modulation index from $2\pi$ to $\pi$, the BER at the receiving end with physical layer decryption by DM2 degrades from error-free to $2.7\times10^{-1}$. When the modulation index is $1.85\pi$, $1.9\pi$, and $1.95\pi$, the BER is $4.3\times10^{-3}$, $2.3\times10^{-3}$, and $1.8\times10^{-4}$, respectively.

When the ideality of the TAI signal with $t_s = 25$ ps is taken into consideration, as the TAI signal bandwidth decreases from 80 GHz to 10 GHz, the BER degrades from error-free to $4.4\times10^{-1}$, as shown in Fig. 12(b). It is important to note that while we are discussing the bandwidth of the TAI signal, the bandwidth of the modulator is also crucial, as it needs to be able to match the bandwidth of the TAI signal. Therefore, in practical applications, the PM is better to have a low half-wave voltage so that the requirement on the TAI signal amplitude will not be that strict. Meanwhile, the larger the bandwidth of the PM, the better. For the TAI signal, it needs to have a high amplitude to apply a $2\pi$ phase change on the optical signal, while it is better to have a very good step characteristic, that is enough signal bandwidth.

Fig. 12(c) shows the BER of the proposed scheme when the bandwidth of the PD is adjusted from 10 GHz to 50 GHz. Correspondingly, the BER at the receiving end in the optical network is improved from $1.3\times10^{-2}$ to error-free after physical layer decryption. Therefore, in practical applications, the PD with greater bandwidth is better to be used, so that the pulses generated after photodetection can have a small distortion.

*3.7 Data transmission under different lengths of the SMF*

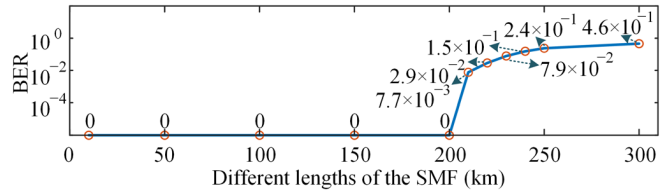

Fig. 13. BER of the system when the SMF with different fiber lengths is used for transmission.

In the previous studies, the length of the SMF is fixed at 10 km. However, in practical optical networks, the distance between the transmitting end and the receiving end is variable and may be longer than 10 km. In this study, data transmission under different lengths of the SMF is further verified under Configuration 3. The length of the SMF is set to 10 km, 50 km, 100 km, 150 km, 200 km, 210 km, 220 km, 230 km, 240 km, 250 km, and 300 km. The two EDFAs in the system, each with a maximum gain of 40 dB, are used to provide sufficient gain to compensate for the SMF loss and ensure a −3.5 dBm optical power launched into the SMF at the transmitting end in the optical network and a −5.5 dBm input optical power of the PD at the receiving end in the optical network. When the length of SMF reaches 220 km, 230 km, 240 km, 250 km, and 300 km, a third EDFA is used after EDFA2 at the receiving end because the maximum gain of the EDFA2 is set to 40 dB. In addition, in this study, the dispersion value of DM2 is corresponding adjusted according to the dispersion introduced by the SMF in the optical network.

As shown in Fig. 13, as the length of SMF increases from 10 km to 200 km, the BER at the receiving end remains error-free, indicating that Configuration 3 demonstrates the capability of error-free transmission over distances of up to 200 km. However, when the transmission distance is extended from 200 km to 300 km, the BER rapidly increases from error-free to $4.6\times10^{-1}$. The primary reason for this is that when the length of the SMF exceeds 200 km, the impact of amplified spontaneous emission noise introduced by the EDFAs at the receiving end gradually increases. It is important to note that the dispersion introduced by the increased fiber length does not lead to a degradation of system performance, as it also contributes to the total amount of dispersion required by the proposed scheme. The loss of the SMF and noise introduced by EDFA are the keys that limit the data transmission performance of the proposed scheme in the optical network.

Under Configuration 3, the system proposed in this work has the capability of error-free transmission up to 200 km. When long-distance error-free transmission is required, data decryption, recovery, re-modulation, and encryption can be performed every 200 km. Furthermore, it is evident that a longer error-free transmission distance of over 200 km can also be directly achieved by reducing the transmission data rate or introducing error correction codes.

## 4. Discussion

### 4.1 Performance improvement based on data randomization

As given and discussed in Section 3.1, before mapping the data matrix to the frequency matrix, a step of randomization is performed on the data matrix. Randomization plays two significant roles in this work: (1) It utilizes random matrices for signal encryption, ensuring that even if the physical encryption is broken, the data remains confidential without knowledge of the specific random matrix used; (2) It randomizes the data to prevent the occurrence of a large number of consecutive "1", thereby enhancing the demodulation performance of the proposed secure data transmission scheme. Why do we need to avoid a large number of consecutive "1"? The explanation is as follows: Since the amplitude of the AWG output signal is constant, the presence of a large number of consecutive "1" can lead to a reduction in the amplitude of each frequency component of the time-varying multi-tone signal, resulting in a decrease in the amplitude of the electrical pulses received at the receiver end, a decrease in signal-to-noise ratio (SNR), and a deterioration of the BER.

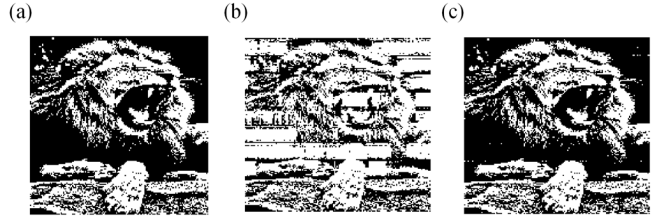

Fig. 14. (a) Binary image transmitted by Alice. (b) Binary image recovered by Bob with physical encryption and decryption and without data randomization. (c) Binary image recovered by Bob with physical encryption and decryption and with data randomization.

The first point is a basic knowledge of secure data transmission. Therefore, it will not be further discussed here. For the second point, we have conducted further analysis and verification. The binary image transmitted by Alice is still the lion as shown in Fig. 14(a). The system configuration remains consistent with the configuration used to obtain the results in Fig. 3, corresponding to a BER of $9.8\times10^{-4}$. As shown in Fig. 14(b), if the data matrix is not randomized and directly transferred to the frequency matrix for modulation, encryption, and transmission, the binary image recovered by Bob after physical decryption will be very blurry with a BER of $2.8\times10^{-1}$. In comparison, when randomization is employed, the blurring in the binary image recovered by Bob after decryption is significantly reduced, which is displayed in Fig. 14(c). Some reversed pixels are due to the BER corresponding to this configuration not being 0.

*4.2 Reducing the risk of information theft using time-varying DM and TAI signal*

In the original fixed asymmetric dispersion matching scheme in Fig. 2, both Alice and Bob in the optical network are constrained to utilize a specific pair of DMs. As given and discussed in Section 3.2, the secrecy rate of the proposed scheme is very close to or equal to 1 when the dispersion of DM2 deviates from the ideal value by 2% or more. This means that the proposed scheme has a very high secrecy rate when the dispersion of DM2 is not exactly matched to the ideal value. In this section, assuming Eve possesses the ability to perform an exhaustive search on unknown dispersion values and uses a complete trial-and-error method, Eve may potentially achieve a dispersion matching with an accuracy better than 2% through numerous attempts with different dispersions, thus enabling the decryption of the physical encryption. To prevent this from happening and further enhance the confidentiality of the physical layer encryption method proposed in this work, the use of time-varying DMs and TAI signals can be further employed to reduce the risk of information theft.

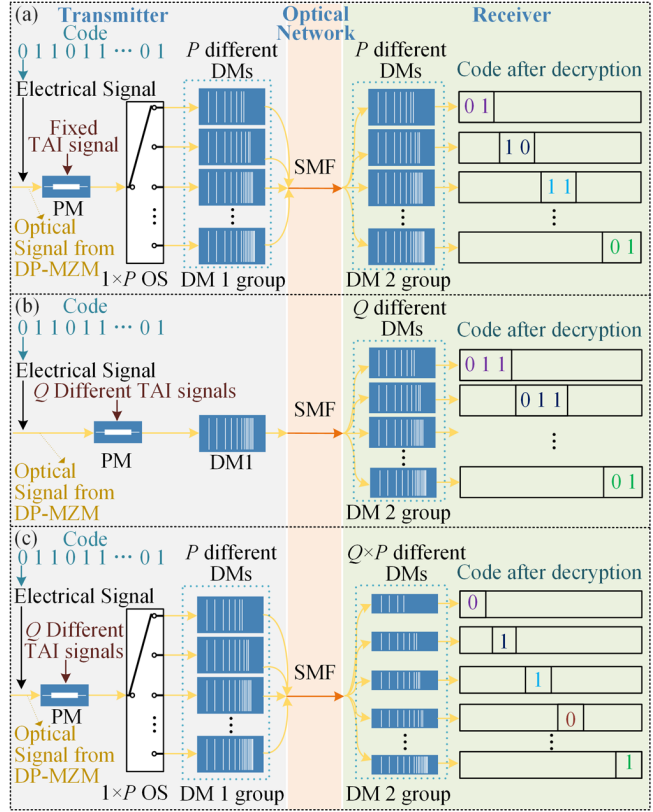

Fig. 15. Schematic diagram of high-speed switching of DM and TAI signal to further reduce the risk of information theft. (a) Switching only the DM, (b) Switching only the TAI signal, (c) Switching both the DM and the TAI signal. PM, phase modulator; DM, dispersive medium; OS, optical switch; SMF, single-mode fiber.

As indicated in Fig. 15(a), at the transmitting end in the optical network, the optical signal carrying the data and after TAI phase modulation from the PM is sent to a $1 \times P$ optical switch (OS). Each output of the OS is connected to an independent DM with a different dispersion. The optical signal input into the OS is controlled by the optical switch and is distributed into $P$ different DMs in the DM1 group in a high-speed and specific order. Then, the outputs from the $P$ different DMs are coupled together as the optical signal to be transmitted. Under these circumstances, the data to be transmitted are encrypted by the $P$ different DMs and multiplexed in the temporal domain. At the receiving end in the optical network, the optical signal after transmission in the SMF is received and processed by the DM2 group. The dispersion of each DM in the DM2 group is matched with that of each DM in the DM1 group according to Eq. (3) and (4). As shown in Fig. 15(a), after being processed by the DM2 group, the data encrypted by different DMs in the DM1 group is recovered at the output of different DMs in the DM2 group. Different from switching DM1 in the transmitting end, another method to reduce the risk of information theft is based on switching the TAI signal, as shown in Fig. 15(b). In this case, the TAI signal applied to the PM changes every period or several periods, so the total dispersion determined by Eq. (4) also changes accordingly. Thus, when DM1 in the transmitting end is fixed, DM2 with different dispersions is required, so a DM2 group is employed at the receiving end. When $Q$ different TAI signals are employed, the DM2 group should theoretically have $Q$ different DMs. Combining the two methods in Fig. 15(a) and (b), the risk of information theft can be further greatly reduced, as shown in Fig. 15(c). In this case, $Q$ different TAI signals and $P$ different DMs are employed at the transmitting end. Thus, at the receiving end, in theory, $Q \times P$ different DMs are required for correctly recovering the original data.

To reduce the risk of information theft using the aforementioned methods in Fig. 15, it is necessary to share switching strategies between Alice and Bob in the optical network. It is crucial to note that this differs from digital decryption algorithms using symmetric keys, which require the sharing of keys: Here, we are dealing with physical encryption, where merely stealing the key is insufficient for decryption; instead, a complete physical system incorporating complex DMs is necessary, significantly increasing the difficulty of decryption even with the knowledge of the switching strategies. Additionally, since only the switching strategies are shared, the specific TAI signal formats, as well as the dispersion values of DMs, do not need to be included in these strategies, and Alice and Bob in the optical network only need to be aware of them locally, which will significantly enhance the reliability of the physical encryption.

*4.3 Advantages of the solution in optical networks*

The proposed secure data transmission scheme has three key advantages when it is applied in optical networks: (1) The encryption and decryption process based on the Talbot effect is realized through physical media, without the need for complex mathematical calculations, thus enabling fast encryption and decryption, which is conducive to improving the overall efficiency of optical networks; (2) Unlike the negative effect of dispersion in traditional optical networks, in the secure data transmission scheme proposed in this work, dispersion is an integral part of the physical encryption and decryption process, therefore, the entire optical network does not require dispersion compensation fibers; (3) The method proposed in Fig. 15 to enhance confidentiality essentially involves performing different encryptions in the time dimension, achieving better secure transmission through more complex encryption methods. However, if the transmitted signals in Fig. 15 are allocated to multiple nodes in the optical network, each node can demodulate the data from different time slots of the transmitted signal through distinct DM2 dispersions, not only ensuring secure transmission but also possessing the potential to integrate with time-division multiplexing optical networks.

## 5. Conclusion

In summary, a novel microwave photonic scheme for secure data transmission in optical networks is proposed based on the temporal Talbot effect in DMs. The binary data to be transmitted in the optical network is converted to a frequency matrix, which is further used to generate a time-varying multi-tone electrical signal carrying the data. The temporal Talbot effect is introduced to noise the optical signal modulated by the time-varying multi-tone electrical signal when the dispersion of the DM at the transmitting end does not match the TAI signal. After fiber transmission through the optical network, at the receiving end, another DM is used in conjunction with the DM at the transmitting end and the optical fiber to fulfill the dispersion value required by the temporal Talbot effect for obtaining compressed pulses, which carry the transmitted data. Therefore, dispersion matching is the key to the proposed approach for secure data transmission in optical fiber. The physical layer security of the proposed scheme and its fiber transmission capability are demonstrated. 8-Gbit/s data is transmitted, encrypted, and decrypted using two DMs and an optical fiber of 10 to 200 km, and error-free transmission is achieved. Many factors that affect the encryption, decryption, and transmission performance of the system have been analyzed. To further reduce the risk of information theft in the proposed scheme, we also give three additional methods that involve switching the TAI signal and the DM. By implementing these methods, the security of data transmission in optical networks will be significantly enhanced. The proposed scheme in this work is expected to be applied as a new physical encryption method in optical networks with higher security requirements.

**Funding.**